\input{aipcheck.tex}

\documentclass[final]
  {aipproc}

\layoutstyle{8x11double}

\SetInternalRegister\hbadness{8000} 

\newcommand\doingARLO[2][]{%
  \ifx\mmref\undefined #1\else #2\fi
}

\begin{document}

\title[Hard X-ray afterglows of short GRBs]
{Hard X-ray afterglows of short GRBs}


\author{Davide Lazzati}{
address={Institute of Astronomy, Madingley Road CB3 0HA Cambridge, U.K.}
}

\iftrue
\author{Enrico Ramirez-Ruiz}{
  address={Institute of Astronomy, Madingley Road CB3 0HA Cambridge, U.K.}
}

\author{Gabriele Ghisellini}{
  address={Osservatorio Astronomico di Brera, via Bianchi 46, 23807 
Merate (LC), Italy}
}
\fi

\copyrightyear  {2001}

\begin{abstract}
We report the discovery of a transient and fading hard X-ray emission
in the BATSE lightcurves of a sample of short $\gamma$-ray bursts.  We
have summed each of the four channel BATSE light curves of 76 short
bursts to uncover the average overall temporal and spectral evolution
of a possible transient signal following the prompt flux.  We found an
excess emission peaking $\sim 30$~s after the prompt one, detectable
for $\approx 100$~s.  The soft power-law spectrum and the
time-evolution of this transient signal suggest that it is produced by
the deceleration of a relativistic expanding source, as predicted by
the afterglow model.
\end{abstract}

\date{\today}

\maketitle

\section{Introduction}

Since their discovery, $\gamma$-ray
bursts (GRBs) have been known predominantly as brief, intense flashes
of high-energy radiation, despite intensive searches for transient
signals at other wavelengths.  Fortunately, the rapid follow-up of
{\it Beppo}SAX 
\citep{Boella97} positions, combined with
ground-based observations, has led to the detection of fading
emission in X-rays \citep{costa97}, optical \citep{vp97} and radio 
\citep{fr97} wavelengths. These afterglows
in turn enabled the measurement of redshifts \citep{met97},
firmly establishing that GRBs are the most luminous known events in
the Universe and involve the highest source expansion velocities.

The detection of afterglows that follow systematically long bursts has
been a major breakthrough in GRB science.
Unfortunately no observation of this kind was possible for short
bursts.  Our physical understanding of their properties was therefore
put in abeyance, waiting for a new satellite better suited for their
prompt localization.

In this paper we show that afterglow emission characterizes also the
class of short bursts. In a comparative analysis of the BATSE
lightcurves of 76 short bursts we detect a hard X-ray fading
signal following the prompt emission with a delay of $\sim 30$ s.  The
spectral and temporal behavior of this emission is consistent with the
one produced by a decelerating blast wave, providing a
direct confirmation of relativistic source expansion. For a more detailed
discussion, see Lazzati, Ramirez-Ruiz and Ghisellini
\citep{laz02}.

\section{Data analysis}

The detection of slowly variable emission in BATSE lightcurves is a
non trivial issue, since BATSE is a non-imaging instrument and
background subtraction can not be easily performed. We selected from
the BATSE GRB catalog a sample of short duration ($T_{90}\le1$~s), high
signal-to-noise ratio, GRB lightcurves with continuous data from
$\sim120$~s before the trigger to $\sim230$~s afterwards. We aligned
all the lightcurves to a common time reference in which the burst
(binned to a time resolution of 64~ms) peaked at $t=0$ and we binned
the lightcurves in time by a factor 250, giving a time resolution of
16.0~s. The time bin [$-8<t<8$~s] containing the prompt emission was
removed and the remaining background modelled with a $4^{\rm th}$
degree polynomial.  The bursts in which this fit yielded a reduced
$\chi^2$ larger than 2 in at least one of the four channels were
discarded. Note that we did not subtract this best fit background
curve from the data. This procedure was used only to reject
lightcurves with very rapid and unpredictable background fluctuations
and is based on the assumption that the excess burst or afterglow
emission is not detectable in a single lightcurve. 
This procedure yielded a final sample
of 76 lightcurves, characterized by an average duration
$\langle{T}_{90}\rangle=0.44$~s and fluence
$\langle{\cal{F}}\rangle=2.6\times10^{-6}$~erg~cm$^{-2}$.

\begin{figure}
  \includegraphics[width=\textwidth]{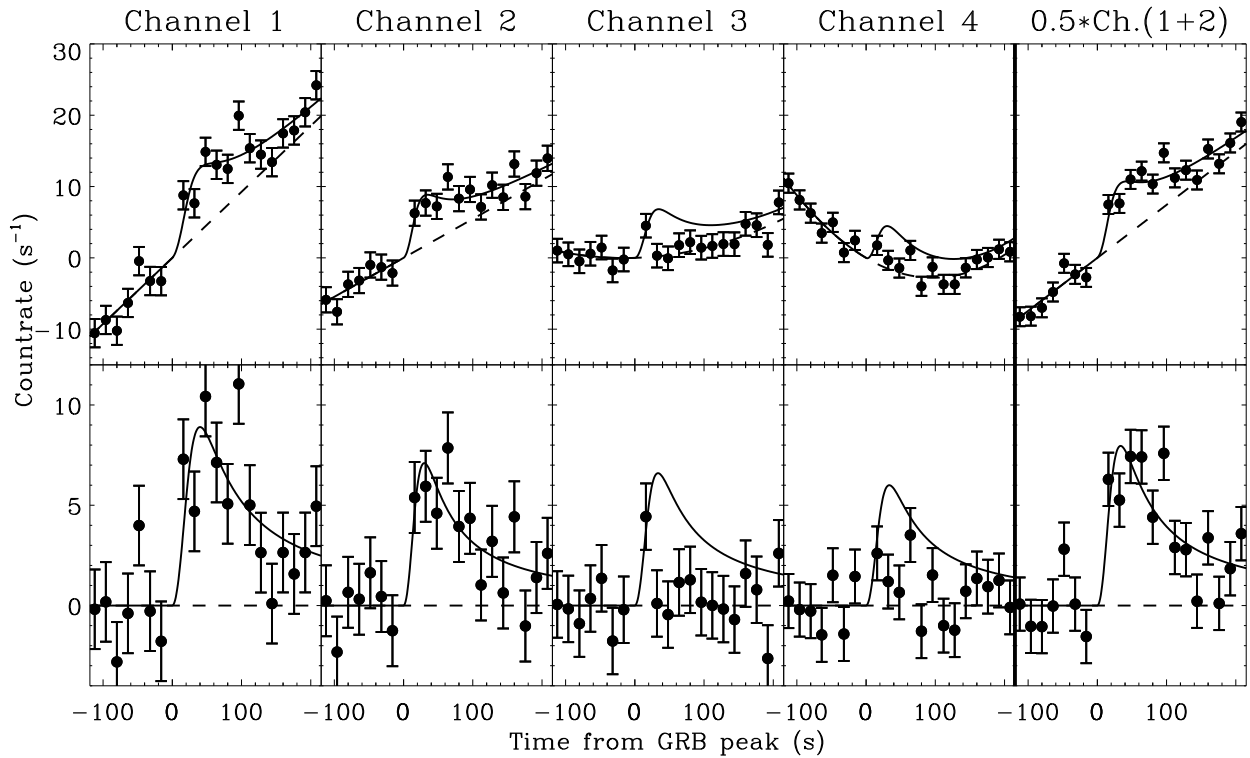}
  \caption{Overall lightcurves in the 4 BATSE channels (from left to
right) of the sample of short bursts (see text). The rightmost panels
show the average signal in the first and second channels. The time
interval of the burst emission has been excluded. The upper panels
show the lightcurves without background subtraction (a constant has
been subtracted in all panels for viewing purposes in order to have
zero counts at $t=0$). The solid line is the best fit background plus
afterglow model (in the channel 3 and 4 panels the $3\sigma$ upper
limit afterglow is shown). The dashed line shows the background
contribution in all channels. The lower panels show the same data and
fit after background subtraction.}
\end{figure}

To search for excess emission following the prompt burst, we added the
selected binned lightcurves in the four channels independently.  The
resulting lightcurves are shown in the upper panels of
Fig. 1 by the solid points. Error bars are computed by
propagating the Poisson uncertainties of the individual lightcurves.

The lightcurves in the third and fourth channels can be successfully
fitted with polynomials. The third (110--325~keV) and fourth
($>325$~keV) channel lightcurves can be fitted with a quadratic model,
yielding $\chi^2/{\rm d.o.f.}=17/18$ and $\chi^2/{\rm d.o.f.}=18.5/17$
respectively.  In the first two channels, a polynomial model alone
does not give a good description of the data. In the first
(25--60~keV) channel, a cubic fit yields $\chi^2/{\rm d.o.f.}=42/16$,
while in the second (60--110~keV) we obtain
$\chi^2/{\rm d.o.f.}=26/16$.  A more accurate modelling of the first
two channels lightcurves can be achieved by allowing for an afterglow
emission following the prompt burst.  We model the afterglow
lightcurve with a smoothly joined broken power-law function:
\begin{equation}
L_A(t) = {{3\,L_A}\over{\left({{t_A}\over{t}}\right)^2 + {{2\,t}\over{t_A}}}};
\;\;\; t>0
\label{eq:ag}
\end{equation}
which rises as $t^2$ up to a maximum $L_A$ that is reached at time
$t_A$ and then decays as $t^{-1}$. Adding this afterglow
component to the fit, we obtain $\chi^2/{\rm d.o.f.}=28.5/16$ and
$14.7/16$ in the first and second channels, respectively. 
The $\chi^2$ variation, according to the F-test, is
significant to the $\sim3.5\sigma$ level in both channels. The fact
that the fit in the first channel is only marginally acceptable should
not surprise. This is because the excess is due to many afterglow
components peaking at different times, and has therefore a more
``symmetric'' shape than Eq.~\ref{eq:ag}. A fully acceptable fit can
be obtained with a different shape of the excess, but we used the
afterglow function for simplicity. By adding together the first two
channels, the afterglow component is significant at the $4.2\sigma$
level. The results of the fit are reported in Tab.~\ref{tab:fit}.

\begin{table}
\begin{tabular}{l|c|c|c}
 & E (keV) & $L_A$ (cts~s$^{-1}$) & $t_A$ (s) \\ \hline \hline
Channel \#1 & [25-60] & $8.9\pm2.6$ & $40\pm16$ \\
Channel \#2 & [60-110] & $7.1\pm2.4$ & $30^{+16}_{-10}$ \\
Channel \#1+2 & [25-110] & $16\pm3.5$ & $33.5^{+24}_{-15}$ \\
Channel \#3 & [110-325] & $<6.6$ & $33.5$ (fixed) \\
Channel \#4 & $>325$ & $<6.0$ & $33.5$ (fixed)
\end{tabular}
\caption{{Fit results. Quoted errors at $90\%$ levels, upper limits
at $3\sigma$ level.}
\label{tab:fit}}
\end{table}

\section{Discussion}

In order to understand whether the excess is residual prompt burst
emission or afterglow emission, we computed its four channel
spectrum. To convert BATSE count rates to fluxes, we computed an
average response matrix for our burst sample by averaging the matrices
of single bursts obtained from the {\tt~discsc\_bfits} and
{\tt~discsc\_drm} datasets. The resulting spectrum is shown in
Fig. 2. The dark points show the spectrum at $t=30$~s
(the peak of the afterglow in the second BATSE channel), which is
consistent with a single power-law $F(\nu)\propto\nu^{-1}$ (black
dashed line). Grey points show the time integrated spectrum (fluences
has been measured with a growth curve technique), consistent with a
steeper power-law $F(\nu)\propto\nu^{-1.5}$ (grey dashed line).  These
power-law spectra are much softer than any observed burst spectrum
(independent of their duration). 
This spectral diversity, together with the fact that a single
power-law does not fit the data, suggest that the emission is not due
to a tail of burst emission but more likely to an early hard X-ray
afterglow. This also confirms earlier predictions that the mechanism
responsible for the afterglow emission is different from that of the
prompt radiation.

An interesting comparison can be made with the early afterglow of
long GRBs.  Connaughton \citep{con00} finds that on average the BATSE
countrate is $\sim150$~cts~s$^{-1}$ at $t=50$~s after the main
event. In our short GRB sample, the countrate at the same time is
$\sim15$~cts~s$^{-1}$.  Since the luminosity of the average early
X-ray afterglow is representative of the total isotropic energy of the
fireball, we can conclude that the isotropic equivalent
energy of the short bursts is on average ten times smaller than that
of the long ones (or that their true energy is the same, but the jet
opening angle is three times larger). Indeed, the $\gamma$-ray fluence
of long bursts is on average ten times larger than that of short
bursts.

In the case of short bursts, the analysis of the afterglow emission is
made easier by the lack of superposition with the prompt burst
flux. For this reason the time and luminosity of the afterglow peak
can be directly measured while in long bursts it had to be inferred
from the shape of the decay law at longer times.
 In our case, however, the lightcurve in
Fig. 1 is the result of the sum of many afterglow
lightcurves, with different peak times and luminosities. For a given
isotropic equivalent energy $E$, afterglows peaking earlier
(with larger $\Gamma$) are expected to be brighter and should dominate
the composite lightcurve. On the other hand, for a given Lorentz
factor $\Gamma$, afterglow peaking earlier (with lower $E$) are
dimmer. The fact that the $\sim35$~s timescale is preserved, suggests
that there are only few very energetic bursts with a large bulk
Lorentz factor.

\begin{figure}
\includegraphics[width=0.48\textwidth]{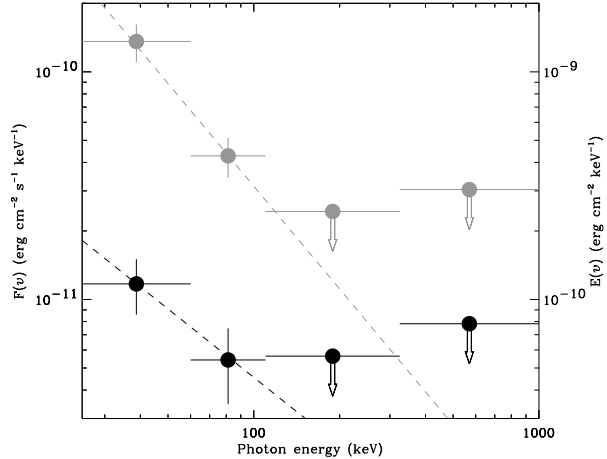}
\caption{Spectrum of the peak afterglow emission. Black dots (and left
vertical axis) show the spectrum at $t=30$~s. Gray dots (right
vertical axis) show the time integrated spectrum as obtained from the
four BATSE channel counts. Error bars are for $90\%$ uncertainties,
while arrows are $3\sigma$ upper limits.}
\end{figure}

We must also remain aware of other possibilities.  For instance, we
may be wrong in assuming that the central object goes dormant after
producing the initial explosion. A sudden burst followed by a slowly
decaying energy input could arise if the newly formed black hole
slowly swallows the orbiting torus around it or if the central object
becomes a rapidly-spinning pulsar rather than a black hole. 
This luminosity may dominate the continuum afterglow at early
times before the blast wave decelerates.  Under this interpretation,
the hard X-ray transient following the prompt emission could be
attributed to the central object itself rather than to a standard
decelerating blast wave. Contrary to what is observed, this emission
should smoothly decay after the main episode, unless this energy is
converted into a relativistic outflow which is in turn converted to
radiation at a larger radius.


\doingARLO[\bibliographystyle{aipproc}]
          {\ifthenelse{\equal{\AIPcitestyleselect}{num}}
             {\bibliographystyle{arlonum}}
             {\bibliographystyle{arlobib}}
          }
\bibliography{davide}

\end{document}